\documentstyle{mn}

\newcommand{\kms}{km\,s$^{-1}$}
\newcommand{\DEG}{^{\circ}}

\title[Masers in R~Aqr and H1$-$36]
      {Discovery of OH and H$_2$O  masers in R~Aquarii and H1$-$36 Arae}

\author[R. J. Ivison, E. R. Seaquist and P. J. Hall]
       {R. J. Ivison$^1$, E. R. Seaquist$^1$ and P. J. Hall$^2$\\ $^1$
Department of Astronomy, University of Toronto, 60 St George Street,
Toronto, Ontario M5S 1A7, Canada\\ $^2$ Australia Telescope National
Facility, CSIRO, P.O.\ Box 76, Epping, NSW 2121, Australia}

\date{Accepted ...
      Received ...
      in original form ...}
\pagerange{000--000}

\begin{document}

\maketitle

 \begin{abstract}
We present the first results from an all-sky maser-line survey of
symbiotic Miras. Interferometric spectral-line observations of R~Aqr
and H1$-$36 have revealed a 22-GHz water maser in the former and
1612-MHz hydroxyl and weak 22-GHz water maser emission from the
latter. H1$-$36 has thus become the first known symbiotic OH/IR
star. We have also detected weak OH line emission from the vicinity of
R~Aqr, but we note that there are small discrepencies between the OH-
and H$_2$O-line velocities and positions. These detections demonstrate
unequivocally that dust can shield some circumstellar hydroxyl and
water molecules from dissociation, even in systems which possess
intense local sources of UV. Finally, we discuss some of the
implications of these observations. The narrow profile of the water
maser in R~Aqr means that there may finally be an opportunity to
determine the system's orbital parameters. We also point out that high
resolution synthesis observations may trace the distribution of dust
in H1$-$36 and R~Aqr, possibly throwing light on the mass-loss process
in symbiotic Miras and placing constraints on the amount of
collimation experienced by UV radiation from their hot, compact
companions.
 \end{abstract}
 \begin{keywords} masers --- binaries: symbiotic --- stars: individual: R
Aquarii and H1$-$36 Arae
 \end{keywords}

\section{Introduction}

The symbiotic Miras, of which R Aquarii and H1$-$36 are well-known
examples, represent around 15 per cent of the total population of
symbiotic stars. They make up the D(usty)-type sub-class (Webster \&
Allen 1975) whose infrared (IR) colours are understood in terms of
dust emission. Their near-IR spectra and light curves reveal the
presence of a Mira-type variable whilst the forests of emission lines
seen in their optical and UV spectra betray hot, compact companions
such as those found in all symbiotic binaries (Whitelock 1987).

Their Mira-type red giant components are believed to be in the final
phases of nuclear burning -- helium shell-burning stars climbing the
asymptotic giant branch. Mass loss can be considerable and is thought
to be driven by radiation pressure on dust grains. In the most
luminous {\it field} Miras the mass loss generates a circumstellar
(CS) shell. Rich in dust, the shell protects against photodissociation
by interstellar ultraviolet (UV) radiation and thereby harbours
molecular species such as OH and H$_2$O.

Schild (1989) has demonstrated that the symbiotic Miras are
practically indiscernable from normal field Miras in respect to the
relationship between pulsational period and mass-loss rate.
Phenomenologically they cluster in the zone where isolated Miras
transform into OH/IR sources, stars which are labelled on the basis of
their copious far-IR emission and their 1612-MHz OH masers. There
has, however, been one important difference between the behaviour of
isolated Miras and that of their cousins in the symbiotic binaries: at
least 75 per cent of field Miras ($>$\,M\,7.5) are associated with OH
maser emission (Sivagnanam, Le Squeren \& Foy 1988) whereas none of
the 16 symbiotic Miras observed at 1612 MHz have shown any strong
evidence of OH emission, nor is there main-line OH emission at 1665 or
1667\,MHz, this despite searches by Wilson \& Barret (1972),
L\'{e}pine \& Nguyen-Q-Rieu (1974), Brocka (1979), Cohen \& Ghigo
(1980) and Norris et al.\ (1984).

Water masers in late-type stars are thought to originate in the same
CS shells as the OH masers. Interferometric observations imply that
the OH masers operate in the outermost regions of the envelope (where
the gas has reached a near-constant velocity) whereas H$_2$O masers
are associated with regions closer to the star. Synthesis observations
of Miras place the H$_2$O masers at radii of $\sim$50 {\small AU}
(Lane et al.\ 1987) and the OH masers at 500---10$^4$ {\small AU}
(Bowers, Johnston \& Spencer 1984). H$_2$O emission typically covers
$\sim$80 per cent of the OH velocity interval which is consistent with
its location inside the OH masing shell (Engels, Schmid-Burgk \&
Walmsley 1988). Until now only 5 symbiotic Miras have been searched
for emission from the H$_2$O $6_{16} \rightarrow 5_{23}$ maser
transition at 22 GHz: upper limits of 0.9---1.7\,Jy were achieved by
Cohen \& Ghigo (1980) for HM~Sge, V1016~Cyg and R~Aqr and by Deguchi,
Nakada \& Forster (1989) for RX~Pup and RR~Tel. R~Aqr was also
searched without success by Dickinson (1976), L\'{e}pine \& Paes de
Barros (1977), L\'{e}pine, Le Squeren \& Scalise (1978) and Bowers \&
Hagen (1984) with upper limits of 1---20\,Jy.

In fact, only two symbiotics have been associated with maser emission
of any kind: H1$-$36 and R~Aqr. R~Aqr supports SiO masers at 43.122
and 42.820\,GHz ($v=1$, $J=1 \rightarrow 0$; $v=2$, $J=1 \rightarrow
0$) and 86.243 GHz ($v=1$, $J=2 \rightarrow 1$) (L\'{e}pine et al.\
1978; Zuckerman 1979; Hall et al.\ 1990a) whilst H1$-$36 has been
detected at the 43.122-GHz transition only (Allen et al.\ 1989).

Unfortunately, despite early promise and regular monitoring, the SiO
emission has been of no use in determining the radial velocity (RV)
curve of the giant components in these systems. The line profiles are
complex and variable (e.g.\ Martinez, Bujarrabal \& Alcolea 1988),
probably due to turbulent motion. SiO masers are thought to lie closer
to the giant ($<5$ stellar radii, McIntosh et al.\ 1989) than either
water or hydroxyl masers (Nyman \& Olofsson 1986), possibly in the
region between the pulsing photosphere and the expanding CS envelope
where grain formation occurs and where, in symbiotic systems, the
molecules may be afforded some protection from degenerate companion
stars.

Here we report the discovery of the first OH and H$_2$O masers
associated with symbiotic Mira systems. We have obtained unambiguous
detections of OH ($J=3/2$, $F=1 \rightarrow 2$) and H$_2$O ($6_{16}
\rightarrow 5_{23}$) emission from H1$-$36 and H$_2$O emission from
R~Aqr. The observations are part of an all-sky maser-line survey of
symbiotic Miras and measurements made at most of the well-known
transitions between 1.6 and 321\,GHz will be reported at a later date
by Seaquist \& Ivison (1994).

\section{Observations and data reduction}

 \begin{table*}
 \caption{Parameters for synthesized beams and emission sources.}
 \begin{tabular}{lllccccc}
\hline
&&&&&&&\\
Source&Array&Line&Noise (mJy&Synthesized beam&\multicolumn{3}{c}{--- Measured
maser parameters ---}\\
&&&bm$^{-1}$\,ch$^{-1}$)&FWHM ($''$), PA&FWHM ($''$), PA&$\alpha$
(B1950)$^{\ddagger}$&$\delta$ (B1950)$^{\ddagger}$\\
&&&&&&&\\
\hline
&&&&&&&\\
R~Aqr&VLA&H$_2$O&11&$1.7\times0.5$, $120\DEG$&$1.8\times0.6$,
$124\DEG$&$23\;41\;14.27\pm0.02$&$-15\;33\;43.3\pm0.2$\\
     &VLA&OH&36&$12.9\times6.7$, $113\DEG$&N/A&N/A&N/A\\
     &ATCA&OH&04&$21.6\times3.4$, $5\DEG$&$23.1\times3.6$,
$7\DEG$&$23\;41\;14.36\pm0.05$&$-15\;33\;40.0\pm4.2$\\
H1$-$36&VLA&H$_2$O&16&$0.72\times0.56$, $148\DEG$&$1.27\times1.02$,
$17\DEG$&$17\;46\;24.19\pm0.03$&$-37\;00\;35.9\pm0.2$\\
       &VLA&OH$^{\star}$&12&$13.7\times10.4$, $173\DEG$&$13.3\times11.0$,
$176\DEG$&$17\;46\;24.17\pm0.14$&$-37\;00\;34.7\pm2.6$\\
       &ATCA&OH$^{\dagger}$&05&$6.6\times5.2$, $154\DEG$&$6.8\times5.6$,
$145\DEG$&$17\;46\;24.24\pm0.05$&$-37\;00\;35.7\pm1.1$\\
&&&&&&&\\
\hline
 \end{tabular}

\noindent
$^{\star}$\,three-channel average. \hspace{0.5in} $^{\dagger}$\,five-channel
average. \hspace{0.5in} $^{\ddagger}$\,Epoch 1993.46 (VLA) and 1993.61 (ATCA).
 \end{table*}

R~Aqr and H1$-$36 were observed during 1993 May 16 using the NRAO Very
Large Array (VLA) in a B/C hybrid configuration.  This configuration,
with a long northern arm, is well-suited to observing sources in the
southern sky. Weather conditions were poor but the most serious
degradation of data quality was due to {\scriptsize GLONASS} satellite
interference at 1612\,MHz, particularly on the shortest baselines.

The bandwidths were 1.5625 and 6.25\,MHz centred at frequencies of
1612.231 and 22235.08\,MHz, respectively, shifted to velocity frames
close to those of the target stars by on-line software (heliocentric
systemic velocities of $-20$ and $-120$\,\kms\ were adopted for R~Aqr
and H1$-$36). At 1612\,MHz the band was split into 256 channels (each
of width 6.104\,kHz = 1.135\,\kms; full coverage $-193$ to
$+153$\,\kms\ for R~Aqr and $-293$ to $+53$\,\kms\ for H1$-$36) and at
22231\,MHz there were 64 channels (each of width 97.656\,kHz =
1.317\,\kms; coverage $-62$ to $+22$\,\kms\ and $-162$ to $-78$\,\kms,
respectively).

The targets were observed at the H$_2$O transition for 50\,min and at
the OH transition for 30\,min. Observations were broken down into
10--15-min scans, interspersed with brief measurements of the nearby
phase calibrators 1748$-$253 and 2351$-$154. The flux densities of the
target stars were tied to those of their phase calibrators which, in
turn, were tied to the flux density of 3C 286 (13.97\,Jy at
1612\,MHz; 2.57\,Jy at 22231\,MHz).

Observations were also performed using the six-element Australia
Telescope Compact Array (ATCA) at Narrabri, NSW during 1993 July 10 in
excellent weather conditions. The Compact Array has dual linear
polarization feeds and our chosen configuration had minimum and
maximum baselines of 153 and 6000\,m. R~Aqr and H1$-$36 were each
observed at a local frequency of 1612.0\,MHz for 120\,min, with
regular phase calibration measurements and a 5-min scan of 1934$-$638
to establish the primary flux density scale. The full bandwidth of
4\,MHz was split into 1024 channels, each of width 3.906\,kHz
(0.726\,\kms). The full velocity coverage for the 1612\,MHz OH line
was thus $-415$ to $+329$\,\kms\ (in the rest frame of the
telescope array).

Finally, we have obtained SiO ($v=1$, $J=1 \rightarrow 0$; $v=2$, $J=1
\rightarrow 0$) spectra of H1$-$36 using the maser superheterodyne
receiver at Parkes (see Hall et al.\ 1990b for details of the observing
procedure).

Data reduction was standard in most respects for both the VLA and ATCA
measurements. The channels were averaged, disgarding those channels
with low sensitivity near the edges of each band (effectively creating
a narrow-band continuum data set). The data were then edited using
standard procedures within the NRAO {\sc aips} package to remove some
of the effects of satellite interference. Gain solutions were
determined for the calibrators and the results were interpolated for
the target sources. Later, after phase calibration, the reduction was
completed by copying the continuum solutions and bad-data flags to the
spectral line data.

Mapping was accomplished using the {\sc mx} routine within {\sc aips}.
Between 750 and 1000 {\sc clean} iterations were employed per channel
and the maps were assembled by {\sc mx} into three-dimensional data
cubes. Beam characteristics and the noise levels in the maps are given
in Table 1. The noise does not compare favourably with the theoretical
limits, primarily because of contamination by satellite interference
at 1612\,MHz.

Finally, the cubes were inspected for signs of emission near the
position and systemic velocity of the target stars. Two slightly
different search methods were employed: in the first, maps were made
of three-channel averages, a method which increases signal-to-noise
for lines wide enough to cover several channels. For very narrow lines
this technique results in a three-fold dilution of the flux whilst
only reducing the background noise by $3^{1/2}$, so in our second
search method we simply mapped the individual channels.

\section{Results}

\subsection{H1$-$36}

 \begin{figure} \vspace{10cm} \caption{ATCA map of the OH 1612-MHz
emission from H1$-$36 made using the average of the 5 channels around
the line peak. Contours levels are set at $-$5, 5, 10, 20, 40, 80 and
160\,mJy.}
 \end{figure}

Line emission was immediately obvious in the ATCA OH data cube of
H1$-$36. The position of the maser was identical, to within the
errors, in each of the five channels in which it could be most clearly
seen. The average position is close to the value obtained during
simultaneous continuum observations at 2378\,MHz (the 0.46-arcsec
positional discrepancy is well within the quadratic sum of their
errors).

There is no evidence that the OH maser-line emission from H1$-$36 was
spatially resolved by the ATCA beam. Source sizes in the individual
channels (and in the five-channel average) were within 10 per cent of
those for the beam and the PAs of source and beam agree to within
10$^{\circ}$. A map of the strongest five channels is shown in Fig.\
1.

 \begin{figure} \vspace{8cm} \caption{Australia Telescope spectrum of
 the 1612-MHz OH maser line in H1$-$36 (1993 July 10). Some spectral
 smearing has occurred due to the lack of on-line frequency
 correction.}
 \end{figure}

Fig.\ 2 shows the ATCA spectrum of H1$-$36 in the region of the OH
emission line, summed over the the whole source. For diagnostic
purposes, we estimate a brightness temperature ($T_{\rm b}$) of $3
\times 10^6$\,K. This assumes that the radius of the emission region is
comparable with those in other Miras (1000\,{\small AU}, Bowers et
al.\ 1983) and that the distance to H1$-$36 is 7.5\,kpc (Whitelock
1987). The emission mechanism is almost certainly that of a maser,
though we note that observations with higher spatial resolution are
necessary before this can be proved beyond doubt.

The close agreement between the continuum and line centroids and
between the observed peak velocity and the systemic velocity of
H1$-$36, determined from SiO and H92$\alpha$ measurements (see Section
4), demonstrate beyond doubt that the maser is associated with the
target star. Positional parameters associated with the detection (as
well as the central velocity and FWHM of the best Gaussian fit to the
line profile) are given in Tables 1 and 2.

Close inspection of the data cube seemed to reveal the presence of a
much fainter line, covering 2 channels, approximately 5 \kms\ to the
red side of the strong peak. Its existence is not borne out by Fig.\
2 where the faint line is buried in the noise.  The spikes near
$V_{\odot}=-114$, $-106$ and $-99$\,\kms\ are due to weak fringes in
the maps --- side effects of satellite interference which could not be
wholly eradicated.

The H1$-$36 OH maser is also clearly visible in the VLA data cube at a
position which agrees with the ATCA result. Again, there is no
evidence that the emission was spatially resolved by the synthesized
VLA beam. The separation between channels in the VLA spectrum is 56
per cent larger than in the ATCA spectrum, but the data never-the-less
serve as a useful variability check on the velocity and strength of
the OH line. Positional and kinematic parameters of the line emission
are given in Tables 1 and 2. The spectrum is shown in Fig.\ 3.

 \begin{figure} \vspace{8cm} \caption{VLA spectrum of the 1612-MHz OH
 maser line in H1$-$36 (1993 May 15).}
 \end{figure}

During the eight weeks between the ATCA and VLA measurements there was
no change in the velocity of the line centre. At first sight, the data
seem to show that the maser was more intense during the later
observation. However, because of the wider VLA channels, it is
probably more meaningful to compare {\em integrated} line fluxes and
these indicate only a marginal brightening. We note that the OH
1612-MHz upper limit ($3\sigma < 132$\,mJy) obtained by Norris et al.\
(1984) implies that the OH line could be variable on a longer
time-scale.

H1$-$36 also has a very weak source of H$_2$O emission approximately
1.2\,arcsec SWW (PA\,$\sim$\,$254^{\circ}$) of the measured 2378-MHz
continuum position. The positional difference is unlikely to be real,
but we note that H1$-$36 is thought to be a widely separated binary
(see Section 4) and there may be grounds for further
investigation. Assuming an emission radius of 50\,{\small AU}, we
calculate that $T_{\rm b}$ is $3 \times 10^6$\,K which is consistent
with maser emission. The peak flux density (integrated over the
source) is at the 10-$\sigma$ level, so we can say nothing definitive
about the morphology of the emission region. Our best fit suggests
that the emission is extended relative to the beam, but we are tempted
to attribute this appearance to poor phase stability because of
unfavourable weather. The H$_2$O line spectrum of H1$-$36 is shown in
Fig.\ 4.

 \begin{figure} \vspace{8cm} \caption{Spectrum of the 22-GHz H$_2$O
 maser line in H1$-$36.}
 \end{figure}

Fig.\ 5 shows the SiO spectra obtained from Parkes. The $v=2$
transition has been detected here for the first time, and is similar
in strength to the $v=1$ line. Both $v=1$ and $v=2$ peaks are at the
2.5-Jy level and have full widths of $\sim$10\,\kms, centred near
$V_{\odot} = -118$\,\kms. The centroid is likely to be slightly
blueward of the systemic velocity. The H1$-$36 profiles in Fig.\ 5 are
reminiscent of those seen in R~Aqr (Allen et al.\ 1989).

 \begin{figure} \vspace{8cm} \caption{Parkes SiO spectra of H1$-$36.
Dashed line: $v=1$, $J=1 \rightarrow 0$;
solid line: $v=2$, $J=1 \rightarrow 0$.}
 \end{figure}

 \begin{table*}
 \caption{Spectral properties of the maser line emission.}
 \begin{tabular}{lllccccc}
\hline
&&&&&&&\\
Source&Array&Line&Observed Peak$^{\star}$&\multicolumn{4}{c}{--- Gaussian fit
parameters ---}\\
&&&Flux Density&Central velocity$^{\ddagger}$&FWHM&Peak flux&Integrated flux\\
&&&(Jy)&$V_{\odot}$ (\kms)&(\kms)&(Jy)&(Jy\,\kms)\\
&&&&&&&\\
\hline
&&&&&&&\\
R~Aqr&VLA&H$_2$O&$0.30\pm0.01$&$-26.61\pm0.16$&$1.69\pm0.21$&$0.30\pm0.03$&$0.54\pm0.09$\\

&ATCA&OH&$0.06\pm0.01$&$-14.04\pm0.23$&$0.98\pm0.33^{\dagger}$&$0.06\pm0.02$&$0.07\pm0.03$\\
H1$-$36&VLA&H$_2$O&$0.23\pm0.02$&$-128.41\pm0.64$&$3.40\pm1.82$&$0.18\pm0.08$&$0.65\pm0.44$\\

&VLA&OH&$0.15\pm0.01$&$-132.66\pm0.39$&$3.76\pm0.89$&$0.13\pm0.03$&$0.54\pm0.17$\\

&ATCA&OH&$0.27\pm0.01$&$-132.47\pm0.15$&$2.85\pm0.34^{\dagger}$&$0.24\pm0.03$&$0.71\pm0.12$\\
&&&&&&&\\
\hline
 \end{tabular}

\noindent
$^{\star}$\,Integrated over whole source in strongest channel.

\noindent
$^{\dagger}$\,No on-line frequency correction. Spectral smearing is
$<$0.61\,\kms\ (R~Aqr) and $<$0.81\,\kms\ (H1$-$36).

\noindent
$^{\ddagger}$ To convert to $V_{\rm lsr}$: $-1.68$\,\kms\ for R~Aqr;
$+6.95$\,\kms\ for H1$-$36.
\end{table*}

\subsection{R~Aquarii}

The 22-GHz data cube of R~Aqr contains a strong, water emission-line
source with a velocity and position which can be unambiguously
associated with R~Aqr. The emission is concentrated inside only three
channels (see Fig.\ 6); the position agrees with that determined using
8440-MHz continuum measurements (obtained within minutes of the
spectral line data) and there is therefore no doubt that the emission
is in the local environment of R~Aqr.

 \begin{figure} \vspace{8cm} \caption{ATCA 1612-MHz spectrum of the
R~Aqr region. We tentatively identify the strongest peak as an OH
maser line at a heliocentric velocity of $-14.04\pm0.23$\,\kms.}
 \end{figure}

The size of our synthesized beam indicates that the brightness
temperature of the radiation $T_{\rm b} > 3 \times 10^3$\,K (for a
distance to R~Aqr of 250\,pc). Alternatively, by assuming an emission
radius of 50\,{\small AU} as we did earlier for H1$-$36, we determine
a higher $T_{\rm b}$ of $6 \times 10^3$\,K. Neither value allows us to
exclude the possibility that the emission has a thermal origin, but
for a thermal mechanism we would expect to see a line width
commensurate with the expansion velocity of the circumstellar
material. The narrow profile observed here leads us to suspect that
the emission mechanism is that of a maser. We also note that at
temperatures much higher than our lower limit, the H$_2$O will begin
to dissociate. Observations with improved spatial resolution are
required in order to determine a more useful lower bound for $T_{\rm
b}$ and to conclusively diagnose the emission mechanism.

The 1612-MHz ATCA data cube of the R~Aqr region contains a weak source
approximately 3.6\,arcsec NE of the water maser
(PA\,$\sim$\,$23^{\circ}$). The distorted shape of the synthesized
beam means that the position of the OH emission is hard to determine
precisely. The errors are large, particularly in declination, so we
cannot rule out the possibility that the OH emission is coincident
with the H$_2$O maser. We are, however, cautious about the reality of
the OH emission, and we also note that the velocity of the line (see
Fig.\ 7) raises doubts about its association with R~Aqr (though it is
not beyond the realms of possibility that we are seeing a masing
region on the far side of the Mira). There is no sign of the emission
in the VLA data cube but this is probably because the noise level is
an order of magnitude higher.

 \begin{figure} \vspace{8cm} \caption{Spectrum of the 22-GHz H$_2$O
 maser line in R~Aqr.}
 \end{figure}

\section{Discussion}

Both R~Aqr and H1$-$36 are sources of intense radio continuum and
optical line emission (Allen 1983; Kaler 1981). It is generally
thought that $10^4$---$10^5$\,K degenerate stars orbit the Mira
components.  The binary periods and separations of these two systems
can only be guessed at, but there have been tentative suggestions of a
44-yr period ($\sim$15\,{\small AU} separation) in R~Aqr (Willson,
Garnavich \& Mattei 1981). In H1$-$36 the separation may be closer to
1000\,{\small AU}, indeed the fact that the reddening towards the Mira
vastly exceeds that towards the ionized nebula lends weight to
arguments that the hot companion lies {\it outside} of the dusty
envelope (Allen 1983).

The case of H1$-$36 is extremely interesting. The distance to this
system lies in the 5---10\,kpc range (Whitelock 1987) and so the OH
and H$_2$O (and SiO) masers must be extremely luminous, in fact if the
H1$-$36 OH maser was located at the distance of R~Aqr (250\,pc) it
would have a peak flux density of between 100 and 500\,Jy. The
estimate by Allen (1983) of 1000\,{\small AU} for the binary
separation ties in well with our maser detections. If the hot star
were embedded in the dusty envelope then we might expect some
disruption of the masing process, either by radiative means ---
dissociation of the molecular gas, for example --- or by dynamically
reducing the path length for amplification (Hall, Wark \& Wright
1987). Because the hot companion is thought to lie outside the CS
shell we can simply accept that the dust is as effective at shielding
molecules from relatively local external UV as it is against
interstellar UV. One possible explanation for the strength of the OH
emission is that the presence of the hot star raises the temperature
of the CS shell and thereby increases the abundance of OH relative to
H$_2$O by dissociation. This is an idea which we shall explore more
carefully in a later paper.

The RV of the H1$-$36 hydroxyl maser is 10---11\,\kms\ blueward of the
H92$\alpha$ radio recombination line detected by Bastian (1992), the
43-GHz SiO line detected by Allen et al.\ (1989) and the value
determined optically by Webster \& Allen (1975). If we are correct in
assuming that the ionized gas is (on the average) stationary with
respect to systemic velocity and that we are seeing emission from
molecular gas flowing towards us at a large radial distance from the
Mira (which seems likely) then this difference represents the terminal
velocity of the Mira wind.

The fact that we do not see a second (redward) peak is unusual in
OH/IR stars, but by no means exceptional (e.g.\ Chengalur et al.\
1993). It was suggested by Allen et al.\ (1989) that SiO masers are
only seen in R~Aqr and H1$-$36 because, unusually, their free-free
emission is {\it optically thin} at millimetre wavelengths thereby
allowing us to see through the ionized portion of their winds to the
masing region within. They argue that other symbiotic Miras could
harbour SiO masers, but their line emission may be hidden by optically
thick, ionized envelopes (see e.g.\ Seaquist, Taylor \& Button
1984). It is conceivable that we only observe a single OH peak from
H1$-$36 because the second masing region is simply hidden behind
line-of-sight ionized gas which is optically thick at 1612\,MHz. If
this is the case then our measurements suggest a lower limit of
$\sim$4 for the optical depth of the ionized gas at 1612\,MHz. Of
course, the single OH peak could be equally well explained by placing
the hot companion on the far side of the shell.

The case of R~Aqr differs from that of H1$-$36 in that we have so far
only detected H$_2$O (and SiO) maser emission, both of which are
thought to originate relatively close to the Mira. The OH maser (which
we have only tentatively identified) is weak relative to normal Miras
at 250\,pc. Perhaps this can be used to place limits on the binary
separation: if the separation is extremely large, as in the case of
H1$-$36, then we would expect to see strong OH emission as the dust
should protect the hydroxyl molecules; a separation of the order
50\,{\small AU} would probably disrupt the water maser (though it may
be located closer to the Mira than is normally the case --- monitoring
the line for turbulent variations would be a good probe of this). We
are still left with many possibilities, but the fact that we see an
anomalously weak OH maser may be interpreted as support for a
separation of 50--1000\,{\small AU} --- a region commonly associated
with OH masers. It is unclear, however, how the hot star could prevent
the masing action throughout the entire shell.

If the peak which we have identified as OH emission is real then its
central velocity (which is 12.6\,\kms\ higher than the water maser)
suggests that it is from the far side of the Mira, where the gas is
flowing away from us. This prompts us to wonder whether the hot star
is currently on the near side of the Mira, inhibiting emission from OH
in that area. Again, the weakness of the OH emission from the far side
of the Mira could be the result of obscuration by intervening ionized
gas.

The most striking feature of the H$_2$O maser spectrum of R~Aqr is the
narrow profile of the emission line. Our data are similar in quality
to the first 86-GHz SiO spectrum of R~Aqr presented by Zuckerman
(1979). It is therefore conceivable that longer, more sensitive
observations will reveal a weak, stable emission plateau spanning
several \kms\ with wildly varying emission peaks as is now known to be
the case with the SiO maser (Martinez et al.\ 1988). We would
therefore be foolish to have too much faith in the
period-determination capabilities of the H$_2$O line, but we
never-the-less point out that the water maser probably lies far from
the turbulent SiO emission region and should therefore be a better
probe of the Mira's RV than the aforementioned SiO line. A large,
single dish (e.g.\ Parkes) could probably monitor the RV of the H$_2$O
maser now that it has been unambiguously detected and associated with
R~Aqr.

\section{Concluding remarks}

The concept of studying symbiotic Miras via their molecular emission
lines is not new. For years it held the promise of an improved
understanding of the circumbinary environment in symbiotics, but
following the largely ineffective maser surveys conducted during the
1970s and 80s there has been a widespread misconception that this
avenue of study is a dead end. It should be noted, however, that there
has never been a maser-line survey which has compared field Miras and
symbiotic Miras at comparable {\em luminosity} levels. Allen et al.\
(1989) showed that symbiotic Miras are probably underluminous in SiO
emission with respect to field Miras, but it has yet to be
demonstrated that the same is true for water and hydroxyl masers. This
fact has not prevented numerous explanations being put forward to
explain the supposed lack of masers in symbiotic stars.

Our maser-line observations of R~Aqr and H1$-$36 have at least shown
that dust {\em can} shield some CS hydroxyl and water molecules from
dissociation, even in systems which possess intense local sources of
UV. The observations described here were made as part of a survey and
inevitably suffered in terms of resolution and sensitivity. In the
future, measurements with much higher spatial (and possibly spectral)
resolution should allow us to determine the morphology of the masing
gas. The molecular material should trace the distribution of dust in
the circumbinary environment and may tell us a great deal about the
process of mass-loss in binary stars and the UV radiation field of the
hot, stellar component in symbiotic systems.

\subsection*{ACKNOWLEDGMENTS}

The authors would like to acknowledge the hospitality the
millimetre-wave group at the California Institute of Technology where
a large fraction of this work was undertaken. We are grateful to Dr
Warwick Wilson for his expert assistance during observations with the
Compact Array. The National Radio Astronomy Observatory (NRAO) is
operated by Associated Universities Inc., under a cooperative
agreement with the National Science Foundation. This work was
supported by an operating grant to ERS from the Natural Sciences and
Engineering Research Council of Canada.

\bsp

\end{document}